\documentclass[aps,reprint,superscriptaddress]{revtex4-2}

\usepackage{graphicx}
\usepackage{braket}
\usepackage{amsmath,amsfonts,amssymb,amstext,amscd,amsthm,bbold}
\usepackage{comment,float}
\usepackage{caption}
\usepackage{xcolor}
\usepackage[colorlinks=true,linkcolor=blue,urlcolor=magenta,citecolor=magenta]{hyperref}
\usepackage{color}
\usepackage{graphicx} 
\usepackage[utf8]{inputenc}

\usepackage[normalem]{ulem}
\DeclareMathOperator{\Tr}{Tr}

\usepackage{adjustbox}

\begin{document}

	\title{Quantum magnetometry using discrete-time quantum walk}

	\author{Kunal Shukla}
	\email{kunalshukla@iisc.ac.in}
	\affiliation{Dept. of Instrumentation \& Applied Physics, Indian Institute of Sciences, C.V. Raman Avenue, Bengaluru 560012, India}
	
	\author{C. M. Chandrashekar}
	\email{chandracm@iisc.ac.in}
	\affiliation{Dept. of Instrumentation \& Applied Physics, Indian Institute of Sciences, C.V. Raman Avenue, Bengaluru 560012, India}
	\affiliation{The Institute of Mathematical Sciences, C. I. T. Campus, Taramani, Chennai 600113, India}
	\affiliation{Homi Bhabha National Institute, Training School Complex, Anushakti Nagar, Mumbai 400094, India}
	
	
\begin{abstract}
Quantum magnetometry uses quantum resources to measure magnetic fields with precision and accuracy that cannot be achieved by its classical counterparts. In this paper, we propose a scheme for quantum magnetometry using discrete-time quantum walk (DTQW) where multi-path interference plays a central role. The dynamics of a spin-half particle implementing DTQW on a one-dimensional lattice gets affected by magnetic fields, and the controlled dynamics of DTQW help in estimating the fields' strength. To gauge the effects of the field, we study the variance of the particle's position probability distribution (PD) and use it to determine the direction of the magnetic field maximally affecting the quantum walk. We then employ statistical tools like quantum Fisher information (QFI) and Fisher information (FI) of the particle's position and spin measurements to assess the system's sensitivity to the magnetic fields. We find that one can use the position and spin measurements to estimate the strengths of the magnetic fields. Calculations for an electron implementing quantum walk of fifty time steps show that the estimate had a root-mean-square error of the order of 0.1 picoTesla. Moreover, the sensitivity of our system can be tuned to measure any desired magnetic field. Our results indicate that the system can be used as a tool for optimal quantum magnetometry.
\end{abstract}	
	

	\maketitle

	\section{\label{sec1}Introduction}

	To detect weak magnetic fields with ultra-high precision is an essential endeavor in diverse areas of science and technology. The leading contender in this field is quantum magnetometry\,\cite{kantsepolskyExploringQuantumSensing2023, hahlMagneticfielddependentStimulatedEmission2022, taylorHighsensitivityDiamondMagnetometer2008, danilinQuantumenhancedMagnetometryPhase2018, vettoliereHighlySensitiveTunable2023}. Its applications stretch from examining neural activities and cardiac signals for diagnosing medical conditions\,\cite{botoMovingMagnetoencephalographyRealworld2018, zhangRecordingBrainActivities2020, bisonRoomTemperature19channel2009, aslamQuantumSensorsBiomedical2023}, detecting magnetic minerals and magnetic anomalies in the mining industries\,\cite{keenanMobileMagneticAnomaly2011} to fundamental studies of magnetism \cite{mccullianBroadbandMultimagnonRelaxometry2020} and several other areas of physics research\,\cite{jacksonkimballProbingFundamentalPhysics2023, xuRecentAdvancesApplications2023}. Quantum magnetometry employs quantum resources like superposition, interference and entanglement to make high-resolution measurements of magnetic fields in a wide range of frequencies. Quantum magnetometers use quantum systems called quantum probes that respond sensitively to the changes in their environment. These probes are usually microscopic, disturb the environment very weakly, and are noninvasive and hence ideal for detection. Quantum magnetometers exploit the inherent fragility of these quantum systems, making them very sensitive to magnetic fields and thus suitable for various applications.
	
On the other hand, we have yet another quantum tool- a quantum analog of the classical random walk called discrete-time quantum walk (DTQW) \cite{venegas-andracaQuantumWalksComprehensive2012, chandrashekarDiscreteTimeQuantumWalk2010a}. DTQW exhibits quantum mechanical properties such as superposition and interference right from the very first step of the walk. Furthermore, by tuning the parameters of the operators evolving the quantum walk, one can control multi-path interference and engineer its dynamics. Consequently, DTQW has been applied to a wide variety of problems. Examples include modeling the dynamics of quantum systems like energy transport in photosynthesis \cite{mohseniEnvironmentassistedQuantumWalks2008, engelEvidenceWavelikeEnergy2007}; simulating quantum phenomena like neutrino oscillation \cite{mallickNeutrinoOscillationsDiscretetime2017a, molfettaQuantumWalksSimulators2016, sahuOpenSystemApproach2024}, localization \cite{joyeDynamicalLocalizationDdimensional2012, chandrashekarDisorderInducedLocalization2013, chandrashekarLocalizedQuantumWalks2015} and topological phase \cite{obuseTopologicalPhasesDelocalization2011, kitagawaExploringTopologicalPhases2010}. Variants of the quantum walk have also been used to simulate Dirac equation \cite{chandrashekarTwocomponentDiraclikeHamiltonian2013, mallickSimulatingDiracHamiltonian2019, mallickDiracCellularAutomaton2016}; modeling relativistic quantum dynamics \cite{strauchRelativisticQuantumWalks2006, chandrashekarRelationshipQuantumWalks2010}; and in designing quantum algorithms \cite {shenviQuantumRandomwalkSearch2003, childsExponentialAlgorithmicSpeedup2003, ambainisCoinsMakeQuantum2004} and quantum computation models \cite{chawlaMultiqubitQuantumComputing2023, lovettUniversalQuantumComputation2010}. Apart from their uses in theoretical modeling and simulation, quantum walks have also been experimentally implemented in multiple physical systems, including NMR \cite{ryanExperimentalImplementationDiscretetime2005}, photonics \cite{broomeDiscreteSinglePhotonQuantum2010,schreiberPhotonsWalkingLine2010, andExperimentalRealizationOnedimensional2014a, peretsRealizationQuantumWalks2008}, cold atoms \cite{dadrasExperimentalRealizationMomentumspace2019a, karskiQuantumWalkPosition2009}, Bose-Einstein condensates \cite{xieTopologicalQuantumWalks2020}, and trapped ions \cite{zahringerRealizationQuantumWalk2010a, xueQuantumWalkLine2009, schmitzQuantumWalkTrapped2009}.
	
	A fairly recent work by Razzoli et al.\cite{razzoliLatticeQuantumMagnetometry2019} connected the idea of quantum magnetometry with the continuous time version of DTQW, known as continuous-time quantum walk (CTQW)\cite{venegas-andracaQuantumWalksComprehensive2012}. CTQW evolves a particle in a Hilbert space exclusively defined by the position sites, depicting its evolution as a continuous function of time. The study focused on a charged spinless particle undergoing CTQW on a finite 2D square lattice in the presence of a locally transverse magnetic field. The paper revealed that position measurements on the ground state of the system can be employed to realize nearly optimal magnetometry. Contrary to CTQW, DTQW evolves the particle in discrete steps of time and in a Hilbert space composed of the tensor product of the position space and the two-dimensional \emph{coin} space. The coin operator associated with the coin space makes DTQW more controllable and better suited for engineering the walk dynamics. 
	
\begin{figure*}[!t]
		\centering
		\adjustbox{width=2.1\columnwidth}{%
			\includegraphics{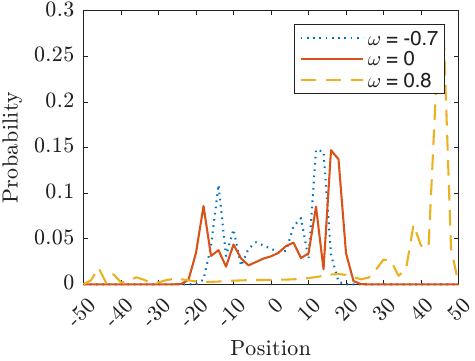}\includegraphics{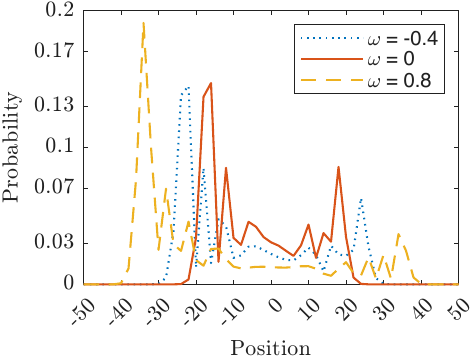}\includegraphics{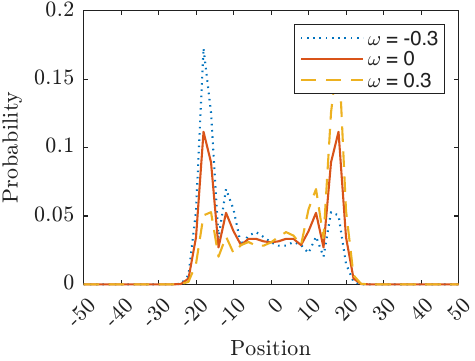}
		}
		\caption{Probability distribution of the position eigenvalues of a spin-half particle after 50 time steps of a bounded DTQW in one dimension. The quantum walk occurs in the presence of magnetic fields of different magnitudes ($\omega$) in the \emph{x}, \emph{y}, and \emph{z}-directions from left to right, respectively. The initial spin states $\ket{0}$, $\ket{1}$, and $\ket{+} = \frac{1}{\sqrt{2}}(\ket{0} + \ket{1})$ respectively are used. In all cases, the coin parameter is set to $3\pi/8$.}
		\label{PPD}
	\end{figure*}

	In this paper, we propose a quantum magnetometry technique that uses DTQW to detect and estimate static magnetic fields homogeneous in space. We assume that their direction is known in advance, and our focus is on detecting the strengths of these magnetic fields. To do so, we implement DTQW on a spin-half particle over a one-dimensional lattice and study the effects of the fields on this system. Our work revolves around two important questions: First, which magnetic field direction affects the DTQW by the largest amount? Second, can we use the fact that magnetic fields affect DTQW (Fig.\ref{PPD}) to detect them and estimate their magnitudes? The second question is naturally followed by finding out the magnitudes and directions which are the most estimatable. 
	
	To answer the first question, we use the variance of the particle's position and spin probability distributions (PD). Our results indicate that based on the form of coin operator used to evolve the quantum walk and the magnitude of field applied, there are some preferred directions of magnetic fields that affect the DTQW of the particle maximally. Moreover, by changing the coin parameter, we can change the direction that has the maximum effect.
	
	Next, to answer the second question, i.e., to quantitatively assess the sensitivity of our system, we use quantum Fisher information (QFI)\cite{parisQuantumEstimationQuantum2009a, giovannettiAdvancesQuantumMetrology2011} as a figure of merit. In addition, we use Fisher information (FI) to find whether measuring the position and spin of the particle provides any information about the external field. The FI we have calculated show that spin and position measurements can indeed be used for optimum magnetometry. The peak value of FI obtained in both cases turns out to be proportional to the square of the total time steps of the DTQW which is in agreement with the bound set by QFI of DTQW \cite{singhQuantumWalkerProbe2019}. Furthermore, by changing the parameters of operators evolving the walk, the peaks of the QFI and FI plots can be desirably shifted. This implies that one can tune the system to be maximally sensitive around any desired magnetic field, making the overall scheme flexible and resilient. Calculations done for an electron undergoing DTQW of only 50 steps show that the magnetometer can estimate the strengths of magnetic fields with the root-mean-square error (RMSE) in the estimate of the order of 0.1 picoTesla.
	
	The paper is structured as follows: In Sec. II, we introduce the system and derive its Hamiltonian for a general static homogeneous magnetic field. We also discuss the evolution of spin-half particle in the DTQW framework. In Sec. III, we discuss the effect of magnetic field over DTQW. We study the variance distribution over magnetic fields of different magnitudes and in different directions. We also analyze which directions of magnetic fields affect the quantum walk maximally. Sec. IV discusses the theoretical framework of quantum estimation theory (QET) used in this work and the main results. Here, we assess the scheme using Fisher information and show why this system has the potential to be a possible magnetometer. Section V closes the paper with some concluding remarks and possible outlooks.
	
	\begin{figure}[h]
	\includegraphics{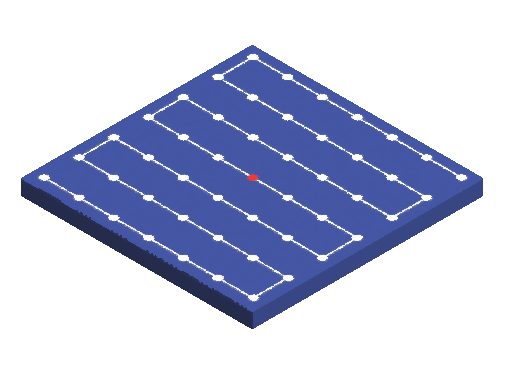}
	\caption{A possible way to design the quantum probe to detect magnetic fields. The red dot is the point $\ket{x=0}$, where a spin-half particle begins the DTQW. The 1D lattice is wrapped in a zigzag manner inside a square to increase the surface area prone to the magnetic field.}
	\label{Probe}
	\end{figure}

	\section{\label{sec2}Probing System}
	
	The quantum system consists of a one-dimensional (1D) lattice (see Fig.\ref{Probe}) made up of $2N + 1$ discrete points marked using integers from $-N$ to $N$. Over this lattice, a spin-half fermion undergoes DTQW. To the whole system, a static homogeneous magnetic field, $\mathbf{B} = B_{0} \mathbf{\hat{n}}$, is applied; we intend to estimate its magnitude.
	
	The Hamiltonian describing a spin-half particle in a magnetic field \cite{shankarSpin1994} is given by
	\begin{equation}
		\hat{\textsf{H}} = -\gamma \hat{\textsf{\textbf{S}}} \cdot \mathbf{B},
	\end{equation}
	where $\hat{\textsf{\textbf{S}}}= (\hat{\textsf{S}}_{x},\hat{\textsf{S}}_{y},\hat{\textsf{S}}_{z})$ is the spin angular momentum operator of the particle and $\gamma \hat{\textsf{\textbf{S}}} = \boldsymbol{\mu}$ is its intrinsic magnetic moment. Since we use a static magnetic field, $\hat{\textsf{H}}$ is time-independent. Using $\hat{\textsf{\textbf{S}}}=\hbar \boldsymbol{\hat{\sigma}}/2$ where $\boldsymbol{\hat{\sigma}} = (\hat{\sigma}_{x}, \hat{\sigma}_{y},\hat{\sigma}_{z})$ and choosing 
	\begin{equation}
		\omega= \gamma B_{0}/2,
	\end{equation} 
	the unitary operator evolving the state of the particle under the influence of the static magnetic field \cite{nakaharaQuantumComputingLinear2008a} can be expressed as
	\begin{equation} \label{eq: unitary_evolution_operator}
		\begin{split}
			\hat{\textsf{U}}(t) &= \exp\left(\frac{-i\hat{\textsf{H}}t}{\hbar}\right) \\
			&= \cos{(\omega t)}\hat{\textsf{I}} + i(\boldsymbol{\hat{\sigma}}\cdot\hat{n}) \sin{(\omega t)}.
		\end{split}
	\end{equation}
	The particle also undergoes DTQW over the 1D lattice, which we will discuss next.
	
	A general DTQW \cite{venegas-andracaQuantumWalksComprehensive2012, chandrashekarDiscreteTimeQuantumWalk2010a} is defined on the Hilbert space $\mathcal{H} = \mathcal{C} \otimes \mathcal{H}_{p}$ where $\mathcal{C} =$ span of $\left\lbrace \ket{0},\ket{1}\right\rbrace $ is called the coin-space of the walker and $\mathcal{H}_{p} =$ span of $\left\lbrace\ket{x}:x \in \mathbb{Z}\right\rbrace$ is called the position-space. In our system, the coin-space is the spin-space of the particle spanned by the eigenvectors of $\hat{\textsf{S}}_{z}$ operator. The position-space, $\mathcal{H}_{p}$, is spanned by states $\left\lbrace \ket{x}\right\rbrace$ where $x= \left\lbrace -N,-N+1,...,N\right\rbrace$. A DTQW is evolved and governed by the following two operators:
	\begin{enumerate}
		\item The coin operator acting on the coin-space $\mathcal{C}$ of the walker given by
		\begin{equation}
			\hat{\textsf{C}}(\tau, \xi, \zeta, \theta)=  e^{i\tau}
			\begin{pmatrix}
				e^{i\xi} \cos(\theta) & e^{i\zeta} \sin(\theta)\\
				-e^{i\zeta} \sin(\theta) & e^{-i\xi} \cos(\theta)
			\end{pmatrix}.  
		\end{equation}
		Throughout our study, we use the coin operator of the form
		\begin{equation} \label{eq: undisturbed_coin}
			\hat{\textsf{C}}(\theta)=  \begin{pmatrix}
				\cos(\theta) & -i\sin(\theta)\\
				-i\sin(\theta) & \cos(\theta)
			\end{pmatrix}.     
		\end{equation}

		\item The conditional unitary shift operator acting on complete Hilbert space ($\mathcal{H} = \mathcal{C} \otimes \mathcal{H}_{p}$) classifies DTQW into two types. The first one is \textit{Unbounded} DTQW, where the walker moves on a position-space of infinite size with the associated shift-operator defined as
		\begin{align}
			\hat{\Lambda}_p^\infty = \sum_{x} \ket{0}\bra{0} \otimes \ket{x-1}\bra{x} + \ket{1}\bra{1} \otimes \ket{x+1}\bra{x}.
			\label{eq: 5}
		\end{align}
		The second type is \textit{Bounded} DTQW (which we use in this paper). It evolves on position-space, $\mathcal{H}_{p}$, with a finite number of sites. The associated position shift-operator bounds the evolution of walk between
		$[-a,a]$ ($a \in \mathbb{Z}$) with boundary condition $\ket{\Psi_{a+1}} = 
		\ket{\Psi_{-a-1}} = 0 $. We define the shift operator as
		\begin{align} 
			\hat{\Lambda}_p^B =& \ket{1}\bra{0} \otimes \ket{-a}\bra{-a} + \sum_{x= -a+1}^{a} \ket{0}\bra{0} \otimes \ket{x-1}\bra{x} \nonumber \\
			+& \sum_{x= -a}^{a-1} \ket{1}\bra{1} \otimes \ket{x+1}\bra{x} + \ket{0}\bra{1} \otimes \ket{a}\bra{a}.
			\label{eq: bounded_shift_operator}
		\end{align}			
	\end{enumerate}
	The operator representing a single step of the DTQW is given by
	\begin{equation}
		\hat{\textsf{W}}=\hat{\Lambda}_p (\hat{\textsf{C}} \otimes \textsf{I}_{p}).
		\label{eq: Step_operator}
	\end{equation}	 
	The quantum walker starts from the position site at the center of the lattice with an initial state of the form $\ket{\Psi(0)}= \ket{s}\otimes \ket{x=0}$.	
	The coin operator generally evolves the coin-state $\ket{s}$ to some superposition of two or more states. The shift operator then shifts those states to the next and/or previous position based on their respective coin states.
	In the next section, we discuss the effects of magnetic fields on DTQW.
		
	 \section{\label{sec3}Effect of Magnetic Field on DTQW}

	 In the presence of a magnetic field, the DTQW can be described in the following way. The particle begins at time $t=0$, with the initial state $\ket{\Psi(0)}$. The operator $\hat{\textsf{W}}$ [Eq.(\ref{eq: Step_operator})] acts on it after regular time-intervals of $\tau$ seconds, acting on $\ket{\Psi(0)}$ for the first time at $t=\tau$ seconds. At any time $t$, the state $\ket{\Psi(t)}$ of the particle is also evolved by the unitary operator $\hat{\textsf{U}}(t)$ [Eq.(\ref{eq: unitary_evolution_operator})] in the presence of a magnetic field. Thus, before the step operator $\hat{\textsf{W}}$ acts for the first time, the particle is in the state $\ket{\Psi'(\tau)}=(\hat{\textsf{U}}(\tau) \otimes \textsf{I}_{p})\ket{\Psi(0)}$. The state of the particle at time $t=\tau$ (when $\hat{\textsf{W}}$ has acted) is thus given by
	 \begin{equation}
	 	\ket{\Psi(\tau)} = \hat{\textsf{W}}(\hat{\textsf{U}}(\tau) \otimes \textsf{I}_p) \ket{\Psi(0)}.
	 \end{equation}
	 Throughout our work, we take the  time  $\tau= 1 s$. However, depending on the situation, one can change the time step value ($\tau$). The state of the particle after $n$ time steps is given by
	 \begin{equation}\label{eq: Psi_n}
	 	\ket{\Psi(n)} = \hat{\textsf{W}}(\hat{\textsf{U}}(1) \otimes \textsf{I}_p) \ket{\Psi(n-1)}.
	 \end{equation}	 
	 
	 \begin{figure*}[t]
	 	\centering
	 	\adjustbox{width=2\columnwidth}{%
	 		\includegraphics{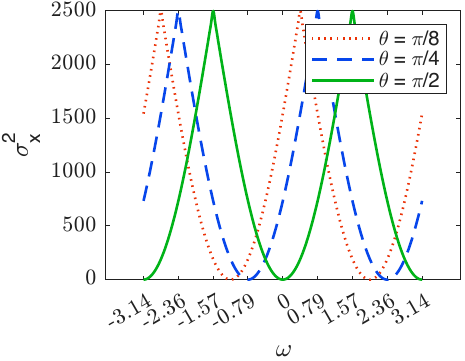}\includegraphics{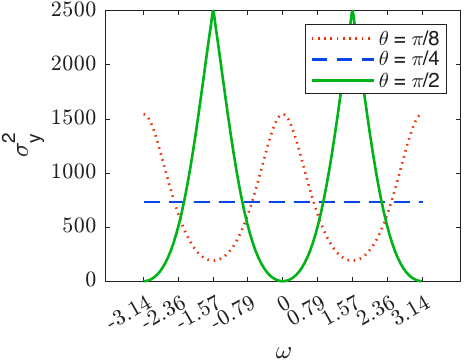}\includegraphics{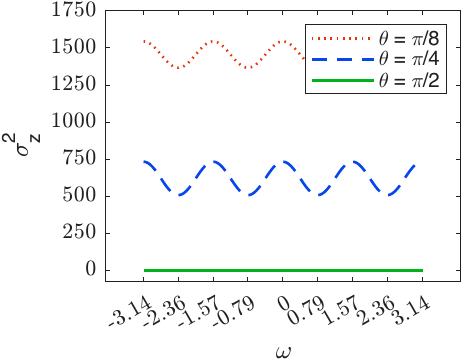}
	 	}
	 	\caption{After 50 steps of the DTQW of a spin-half particle, the variance of the position probability distribution is plotted against various magnitudes ($\omega$) of magnetic fields. Magnetic fields in the \emph{x}, \emph{y}, and \emph{z}-directions, from left to right, are respectively used. Initial spin state $\ket{+}$ is used in all plots.}
	 	\label{variance_dist}
	 \end{figure*}
	 Figure \ref{PPD} shows the probability distribution (PD) of position measurements after 50 steps of a bounded DTQW. The walks evolve in the presence of magnetic fields of different magnitudes in the x,y, and z directions, respectively. Magnetic fields in x and y directions (B$\mathbf{\hat{i}}$ and B$\mathbf{\hat{j}}$) can change the spread of the walk. For magnetic fields in the z-direction, we observe that the spread of the walk is not much affected; however, the PD gets skewed. Furthermore, the PD is positively (negatively) skewed when the field direction is towards the positive (negative) \emph{z}-direction. We will discuss the changes to the spread and the skewness of the walk in some detail now.
	 
	 \subsection{Effect of magnetic field on the position probability distribution of DTQW}	 
	 
	  In this section, we aim to quantify the effect of a magnetic field on the position PD of the quantum walk. To carry out this task, we use the variance of the position PDs: Pick one direction and study how the variance is distributed over different magnitudes of the field in that direction. Additionally, we use the difference of the variance:
	  \begin{equation} \label{eq: diff_variance}
	  	\Delta \sigma^2(\mathbf{B}) = |\sigma^2(\mathbf{B})-\sigma^2(\mathbf{B} = \mathbf{0})|,
	  \end{equation} 
	  as a measure of how strongly a field affects the quantum walk. This helps us identify the direction(s) of magnetic field(s), which affects the quantum walk maximally. We begin with magnetic field in an arbitrary direction.
	  
	  For a static magnetic field, homogeneous in space, $\mathbf{B} = \text{B}_{0} \mathbf{\hat{n}}$, the evolution operator, $\hat{\textsf{U}}_{\mathbf{B}}(1)$ [Eq.(\ref{eq: unitary_evolution_operator})], takes the following form:	
	  \begin{equation}\label{eq: Evolving_Unitary}
	  	\hat{\textsf{U}}_{\mathbf{B}}(1)=
	  	\begin{pmatrix}
	  		\cos(\omega) +i n_z \sin(\omega) & \sin(\omega)(n_y + i n_x)\\
	  		\sin(\omega)(-n_y + i n_x) & \cos(\omega) -i n_z \sin(\omega)
	  	\end{pmatrix},   
	  \end{equation}
	  where $\mathbf{\hat{n}} = (n_x, n_y, n_z)$ is an arbitrary direction. Let
	  \begin{equation}\label{eq: Coin_in_B1}
	  	\hat{\textsf{C}}_{\mathbf{B}} = (\hat{\textsf{C}} \otimes \textsf{I}_{p}) (\hat{\textsf{U}}_{\mathbf{B}}(1) \otimes \textsf{I}_{p}),
	  \end{equation}
	  substituting Eq.(\ref{eq: Coin_in_B1}) into Eq.(\ref{eq: Psi_n}) and using Eq.(\ref{eq: Step_operator}), we can write the state of the walker after t steps of DTQW in the presence of $\mathbf{B}$ as
	  \begin{equation} \label{eq: Psi_B}
	  	\ket{\Psi_{\mathbf{B}}(t)}= \hat{\Lambda}_p \hat{\textsf{C}}_{\mathbf{B}} \ket{\Psi_{\mathbf{B}}(t-1)}.
	  \end{equation}
	  Magnetic fields affect the coin operator of DTQW. Based on the direction and magnitude of the field, $ \hat{\textsf{C}}_{\mathbf{B}}$ takes different forms and evolves the walk in different ways.
	  
	  We call the direction along the spin $\ket{0}$, satisfying $\hat{\textsf{S}}_{z}\ket{0} = \hbar/2 \ket{0}$, the positive \emph{z}-direction. The coin operator \textsf{C}$(\theta)$ [Eq.(\ref{eq: undisturbed_coin})], undisturbed by the magnetic field,  is of the form $\exp{(-i \theta \sigma_{x})}$. Hence, it rotates the spin by angle $2\theta$ about the positive \emph{x}-direction, ascertaining the \emph{x}-direction for us. The positive \emph{y}-direction is chosen perpendicular to both \emph{x} and \emph{z}, following the right-handed coordinate system.
	  
	  Figure \ref{variance_dist} illustrates how variance is distributed over different magnitudes ($\omega = \gamma B_0 /2$) of magnetic fields in the \emph{x}, \emph{y} and the \emph{z} direction, respectively. Rightmost panels in Fig.\ref{PPD} and Fig.\ref{variance_dist} show that the magnetic field in the \emph{z}-direction, B$\mathbf{\hat{k}}$, changes the skewness of the position PD and minimally affects its variance. This behavior is attributed to the resulting form of the coin operator $\hat{\textsf{C}}_{\mathbf{B}}$ when $\mathbf{B} = B_0 \mathbf{\hat{k}}$ is applied to the system, given by
	   \begin{equation} \label{eq: coinz}
	  	\hat{\textsf{C}}_{B_0 \mathbf{\hat{k}}}=  \begin{pmatrix}
	  		\cos(\theta) \exp(i \omega) & -i\sin(\theta) \exp(-i \omega)\\
	  		-i\sin(\theta) \exp(i \omega) & \cos(\theta) \exp(-i \omega)
	  	\end{pmatrix}.    
	  \end{equation}
	  Direct calculations with $\theta = \pi/2$ starting from the walker's initial state $\ket{+} \otimes \ket{x=0}$ reveal that the walker lands in the state `$-\ket{+} \otimes \ket{x=0}$' after just two steps of DTQW. This explains the flat `zero' line for $\theta = \pi/2$ in the rightmost panel of Fig.\ref{variance_dist}. Notably, this nature remains independent of $\omega$, rendering it impossible to estimate the value of $\omega$ using $\hat{\textsf{C}}_{B_0 \mathbf{\hat{k}}}$ with $\theta = \pi/2$ (see the $H_z$ plots in Fig.\ref{QFI}).
	  The skewness of the position PD can also be seen right from the first step of DTQW in presence of B$\mathbf{\hat{k}}$. Using Eq.(\ref{eq: coinz}), starting from initial spin state $\ket{+}$ if we calculate the probability of measuring $x=-1$ and $x=1$ after the first step of the walk, we get
	  \begin{equation}
	  	P(x=\pm1) = \frac{1 \pm\sin(2 \omega) \sin(2 \theta)}{2},
	  \end{equation}
	  showing that B$\mathbf{\hat{k}}$ (-B$\mathbf{\hat{k}}$) skews the position PD towards positive (negative) direction in the lattice right from the first step of the walk. In contrast, using $\hat{\textsf{C}}(\theta)$ [Eq.(\ref{eq: undisturbed_coin})], the coin operator in the absence of $\mathbf{B}$, results in a symmetric PD with $P(x=\pm1)$ equal to half.

	  The field along the \emph{x}-direction, B$\mathbf{\hat{i}}$, is unique in the sense that the variance plot, $\sigma_{x}^2(\omega)$, can be shifted by any desired amount ($\Delta \omega$) by changing the coin parameter, $\theta$, by the same amount. In fact, from numerical results we found the analytical expression governing the variance distribution:
	  \begin{equation}
	  	\sigma^{2}_{x} = -T^2 |\sin(\omega-\theta)| + T^2,
	  \end{equation}	
	  where $T$ corresponds to the total time steps the walker takes in the DTQW. This equation holds true only when the total time steps of the walker is less than or equal to the boundary point of the 1D lattice ($T\le a$), and initial spin state, $\ket{+} = 1/\sqrt{2}(\ket{0}+\ket{1})$, is used. However, even for cases when initial spin state $\ket{s} \ne \ket{+}$ and $T\ge a$, the variance ($\sigma^{2}_{x}$) shows similar nature and shifts by an amount equal to the change in coin parameter ($\Delta \theta$). This happens due to the nature of coin operator, $\hat{\textsf{C}}_{\mathbf{B}}$, when $\mathbf{B} = B_0 \mathbf{\hat{i}}$. The operator takes the following form: 
	  \begin{equation} \label{eq: coinx}
	  	\hat{\textsf{C}}_{B_0 \mathbf{\hat{i}}}=  \begin{pmatrix}
	  		\cos(\theta - \omega) & -i\sin(\theta - \omega)\\
	  		-i\sin(\theta - \omega) & \cos(\theta - \omega)
	  	\end{pmatrix},    
	  \end{equation}
	  explaining the \emph{shift} observed exclusively for magnetic fields in the \emph{x}-direction.
	  
	  Variance, $\sigma^2_x(\omega)$, for cases when $\ket{s} \ne \ket{+}$ and $T\ge a$, must be solutions of the wave-like differential equation:
	  \begin{equation}
	  	\frac{\partial^2 u}{\partial \omega^2}+\frac{\partial^2 u}{\partial \theta^2} = 0,
	  \end{equation}	
	  where $u=\sigma_{x}^2(\omega, \theta)$. We assert this because of the \emph{shift} observed in all the cases when fields in the \emph{x}-direction affect the DTQW.
	  
	  In the presence B$\mathbf{\hat{j}}$, the variance distribution, $\sigma_{y}^2(\omega)$, can only be shifted by multiples of $\pi/2$. Whereas, $\sigma_{z}^2(\omega)$, due to B$\mathbf{\hat{k}}$, cannot be shifted (see Fig.\ref{variance_dist}). At    $\theta =\pi/4$, the variance distribution, $\sigma_{y}^2$ due to B$\mathbf{\hat{j}}$, becomes constant implying that when coin parameter is set to $\theta =\pi/4$, B$\mathbf{\hat{j}}$ does not affect the position PD of the particle. 
	  
	  It is important to emphasize that static homogeneous magnetic fields alter the coin operator in DTQW [Eq.(\ref{eq: Coin_in_B1})]. To gain a deeper understanding of the walk's dynamics in the presence of any magnetic field, we can analyze the impact of the effective coin operator, similar to our examination of B$\mathbf{\hat{k}}$ [Eq. \ref{eq: coinz}] and B$\mathbf{\hat{i}}$ [Eq.(\ref{eq: coinx})]. The availability of \emph{no-shift}, \emph{limited-shift}, and \emph{complete-shift} options for B$\mathbf{\hat{k}}$, B$\mathbf{\hat{j}}$, and B$\mathbf{\hat{i}}$ respectively is attributed to the nature of the obtained $\hat{\textsf{C}}_{\mathbf{B}}$ for these fields. Specifically, for B$\mathbf{\hat{k}}$, $\omega$ only appears as a phase [Eq.(\ref{eq: coinz})]; whereas for B$\mathbf{\hat{j}}$ and B$\mathbf{\hat{i}}$, $\omega$ appears in the form $f(\omega - \theta)$ (allowing the \emph{shift}) within the effective coin operator.
	  
	  \begin{figure}[h]
	  	\includegraphics[width=1\columnwidth]{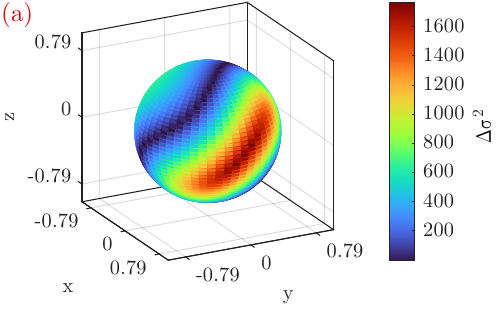}
	  	\includegraphics[width=0.49\columnwidth]{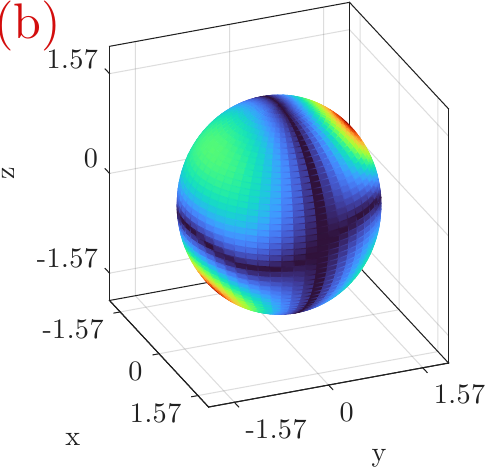}
	  	\includegraphics[width=0.49\columnwidth]{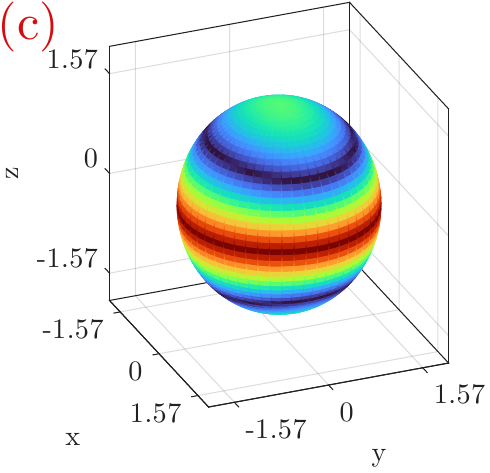}
	  	\includegraphics[width=0.49\columnwidth]{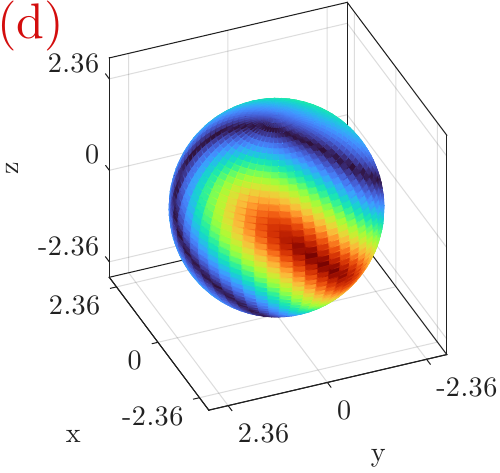}
	  	\includegraphics[width=0.49\columnwidth]{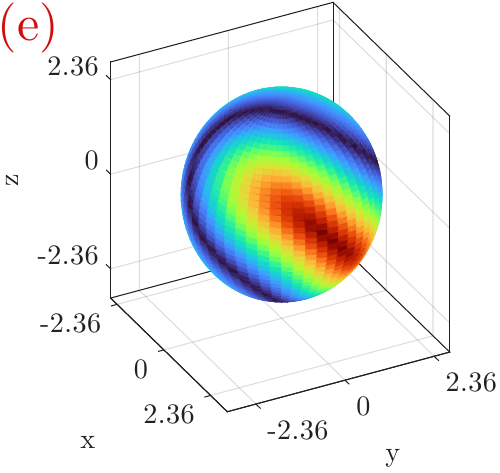}
	  	
	  	\caption{The variance difference, $\Delta \sigma^2(\mathbf{B})$ [Eq.(\ref{eq: diff_variance})], is plotted for magnetic field vectors ($\omega\mathbf{\hat{n}}$) mapped to position vectors of the points on the spheres. In plots (a), (b), and (d) coin parameter, $\theta = \pi/4$ is used for the field of magnitudes, $\omega = \pi/4, \pi/2$, and $3\pi/4$, respectively. In plots (c) and (e), coin parameters $\theta=\omega = \pi/2$ and $\theta=\omega=3\pi/4$, respectively, are used.}
	  	\label{spherical_dist}
	  \end{figure}
	  
	  To determine the direction of the field maximally affecting the quantum walk, we use Eq.(\ref{eq: diff_variance}) and plot the values on a sphere. Position vectors of the points on the sphere are mapped to the vectors $\omega\mathbf{\hat{n}}$. Hence, the radius of the sphere reflects the magnetic field strength, its spacial location depicts the direction of the field, and the color bar shows the value of $\Delta \sigma^2$ calculated at that field. As the field strength is changed, the direction maximally affecting the position PD also changes (see plots (a), (b), and (d) in Fig.\ref{spherical_dist}). Due to the \emph{shift} available for fields in the \emph{x}-direction,  we can make the \emph{x}-direction as the maximally affecting direction for any magnitude of the field being used (depicted in plots (c) and (e) in Fig.\ref{spherical_dist}). It can be done by simply putting $\omega = \theta$ in Eq.(\ref{eq: coinx}). This turns $\hat{\textsf{C}}_{B_0 \mathbf{\hat{i}}}$ into an identity operator allowing the shift operator to take the position state(s) of the walker to the end(s) of the lattice maximizing the variance.
	  
	  It is, however, important to note: To make a magnetic field have the maximum impact on the position PD, the changes done to the coin operator may not allow the field to become easily estimatable. In particular, turning the coin operator into identity reduces the estimatability by inhibiting multi-path quantum interference otherwise available in DTQW dynamics (see minima in Fig.\ref{QFI}). In the following, we delve into the tools and techniques employed to assess and increase the estimatability of magnetic fields within our proposed scheme.
 
	\section{Quantum Estimation in Discrete-time quantum walk}
	This section is divided into two parts. The first part discusses the theoretical tools for estimating an unknown parameter. We review concepts such as Fisher information (FI), quantum Fisher information (QFI), and Cramer-Rao bound and discuss how we have used them in our work. In the second part, we state the main results of this paper and assess the performance of our system in estimating $B_0$.
	
	\subsection{Statistical tools for parameter estimation}
	The problem of quantum parameter estimation \cite{parisQuantumEstimationQuantum2009} is described as follows: We have a quantum system that interacts with some physical quantity with some unknown parameter, say $\lambda$. Here, the unknown parameter $\lambda = B_0$, $B_0$ being the magnitude of the static homogeneous magnetic field we aim to detect. The system's state depends on the unknown parameter and is defined by the density matrix $\rho_{\lambda}$. Our aim is to estimate the value of the unknown parameter ($\lambda$). 
	
	One usually performs measurements on an ensemble of copies of the system to estimate $\lambda$. The results form a random sample: $\{x_1, x_2, ..., x_M\}$. The unknown parameter, $\lambda$, is then estimated by some data-processing of the sample using a function of the form: $\hat{\lambda} = f(x_1, x_2, ..., x_M)$, called the estimator of $\lambda$. 
	
	The simplest way to characterize the quality of the estimate is by calculating the root-mean-square error (RMSE) \cite{giovannettiAdvancesQuantumMetrology2011} of the estimated value ($\hat{\lambda}$) from the true value ($\lambda$) of the parameter. The RMSE is given by
	\begin{equation}
		\delta \hat{\lambda}_M = \sqrt{\langle (\hat{\lambda} - \lambda)^2 \rangle} = \sqrt{\text{Var}(\hat{\lambda})  + [\,b(\hat{\lambda})]\,^2},
	\end{equation}
	where $M$ is the number of measurements made, $\text{Var}(\hat{\lambda}) = \langle (\hat{\lambda} - \langle \hat{\lambda} \rangle)^2 \rangle$ is the variance and $b(\hat{\lambda}) = \langle \hat{\lambda} \rangle - \lambda$ is the bias of the estimator calculated over all possible random samples. When the mean of the estimator, $\langle \hat{\lambda} \rangle$, is equal to the true value of the parameter, i.e., when the estimator is \emph{unbiased}, the RMSE becomes equal to the square root of the estimator’s variance. Smaller the RMSE, better is the estimation accuracy.  Assuming that the estimator is asymptotically locally unbiased, the lower bound on the RMSE is given by the Cramer-Rao inequality:
	\begin{figure*}[!t]
		\centering
		\adjustbox{width=2\columnwidth}{\includegraphics{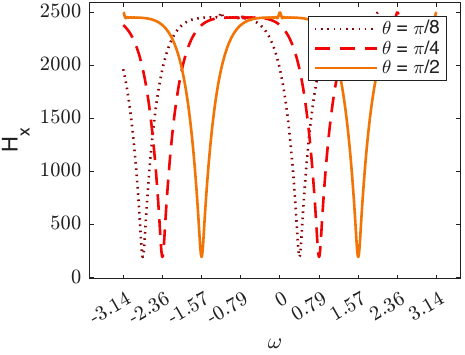}\includegraphics{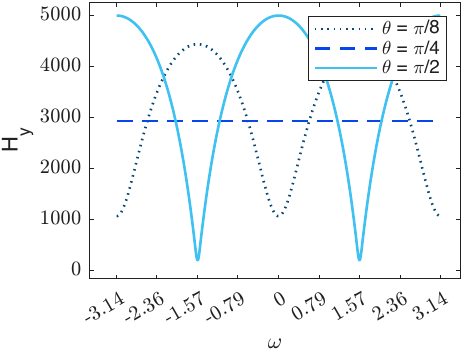}\includegraphics{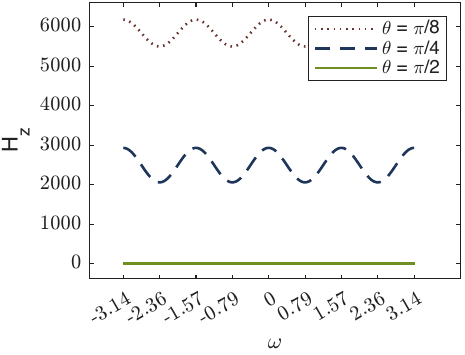}}
		\adjustbox{width=2\columnwidth}{\includegraphics{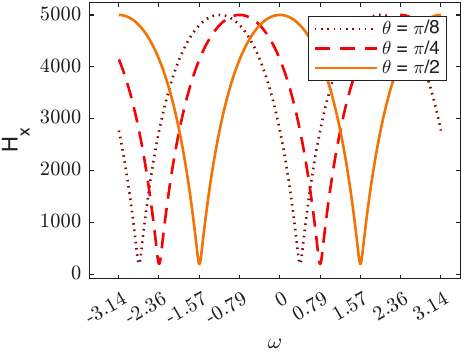}\includegraphics{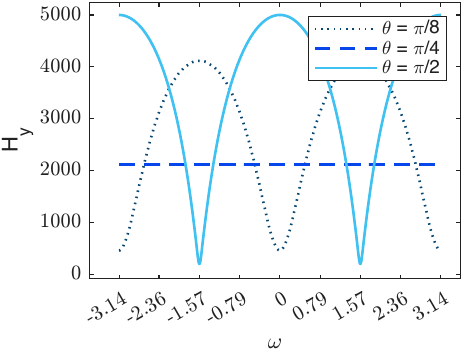}\includegraphics{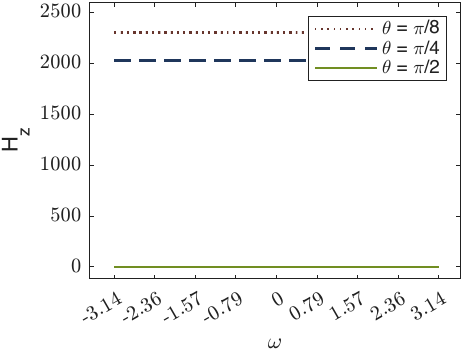}}
		\caption{Distribution of quantum Fisher information over field magnitudes ($\omega$) after 50 steps of DTQW. Initial spin state, $\ket{+} = 1/\sqrt{2}(\ket{0}+\ket{1})$, is used for the DTQW in all three plots at the top. Whereas particle with initial spin state, $\ket{0}$ or $\ket{1}$, is used for three plots at the bottom. From left to right, magnetic fields in the \emph{x}, \emph{y}, and \emph{z}-direction are used.}
		\label{QFI}
	\end{figure*}
	
	\begin{equation} \label{eq: Cramer Rao}
		\delta \hat{\lambda}_M = \sqrt{\text{Var}(\hat{\lambda})} \ge \frac{1}{\sqrt{M F(\lambda)}},
	\end{equation}
	where the quantity $F(\lambda)$ is called the Fisher information. It is defined as
	\begin{equation} 
		F(\lambda) = \int \mathop{dx} p(x|\lambda) [\partial_{\lambda} \ln p(x|\lambda)]^2, 
	\end{equation}
	where $p(x|\lambda)$ is the conditional probability of getting an outcome $x$ when the value of parameter is $\lambda$.
	
	In the quantum setting, the probability of getting $x_i$ as the result of a measurement is given by: $p(x_i|\lambda) = \Tr[\Pi_i \rho_{\lambda}]$, where the set of operators $\{\Pi_i\}$ satisfying $\int \mathop{dx_i} \Pi_i = \mathbb{1}$, form a positive operator-valued measure (POVM). When $\rho_{\lambda}$ is a pure state, $\Pi_i$ is the projection operator ($\ket{x_i}\bra{x_i}$) corresponding to eigenvalue $x_i$. When the eigenspectrum of an observable is discrete, the Fisher information takes the form:
	\begin{equation} \label{eq: FI}
		F(\lambda) = \sum_{i} \frac{(\partial_{\lambda} p(x_i|\lambda))^2}{p(x_i|\lambda)}
		= \sum_{i} \frac{(\partial_{\lambda} \Tr[\Pi_i \rho_{\lambda}])^2}{\Tr[\Pi_i \rho_{\lambda}]}.
	\end{equation}
	The above formula gives the Fisher information provided by measurements related to POVM $\{\Pi_i\}$ on a system described by the density matrix $\rho_{\lambda}$. For different POVMs, one gets different values of FI. This immediately demands finding a POVM which maximizes Fisher information for a given density matrix. We then get the so-called quantum Fisher information (QFI) that theoretically gives the maximum information a system can provide about an unknown parameter. Note that the POVM corresponding to QFI may not always translate to a physically measurable observable.
	
	To calculate QFI, a quantity called symmetric logarithmic derivative (SLD) can be introduced satisfying the relation:
	\begin{equation}
		\frac{1}{2}(L_\lambda \rho_{\lambda}+ \rho_{\lambda} L_\lambda) = \frac{\partial \rho_{\lambda}}{\partial \lambda}.
	\end{equation}
	Using SLD, quantum Fisher information ($H(\lambda)$) can be written in the following form:
	\begin{equation} \label{eq: QFI}
		F(\lambda) \le H(\lambda) = \Tr[\rho_{\lambda} L_\lambda^2],
	\end{equation}	
	where $F(\lambda)$ is the FI. The SLD can be written in a simplified form when one works with pure states. For pure states, we have $\rho_{\lambda}^2 = \rho_{\lambda}$ and hence, $\partial_{\lambda} \rho_{\lambda} = (\partial_{\lambda} \rho_{\lambda}) \rho_{\lambda} + \rho_{\lambda} (\partial_{\lambda} \rho_{\lambda})$. Using the last equation, SLD reduces to: 
	\begin{equation}\label{eq: SLD}
		L_\lambda = 2 (\partial_{\lambda} \rho_{\lambda}).
	\end{equation}
	The inequality in Eq.(\ref{eq: QFI}) indicates that to assess the performance of measurements corresponding to a POVM in estimating an unknown parameter, we can use the ratio:
	\begin{equation}
		R = \frac{F(\lambda)}{H(\lambda)} \le 1.
	\end{equation} 
	The closer the ratio is to one, the higher the efficiency of the POVM. In addition, assuming an unbiased efficient estimator saturating Cramer-Rao Bound [Eq.(\ref{eq: Cramer Rao})] is available, we can use Eq.(\ref{eq: Cramer Rao}) and employ the RMSE to gauge the accuracy and the precision of our estimation with respect to the true value of the unknown parameter.	

	\subsection{QFI of the system and FI of particle's position and spin measurements after DTQW in the presence of a magnetic field}
	
	We begin by performing some calculations needed to compute the QFI provided by our system. As mentioned before, we assume that the direction ($\mathbf{\hat{n}}$) of the magnetic field is already known, and the unknown parameter ($\lambda$) we want to estimate is its magnitude, $\text{B}_{0}$. The state of the walker after $t$ steps of DTQW in the presence of $\mathbf{B}$ is given by Eq.(\ref{eq: Psi_B}). The partial derivative of the density matrix, $\rho_{\mathbf{B}} = \ket{\Psi_{\mathbf{B}}}\bra{\Psi_{\mathbf{B}}}$, is then given by:
\begin{equation}\label{eq: d_rho}
	\partial_{B_0} \rho_{\mathbf{B}} = \ket{\partial_{B_0}\Psi_{\mathbf{B}}}\bra{\Psi_{\mathbf{B}}} + \ket{\Psi_{\mathbf{B}}}\bra{\partial_{B_0}\Psi_{\mathbf{B}}},
\end{equation}
where $\ket{\partial_{B_0}\Psi_{\mathbf{B}}}$ at some time step (t) can be written using Eq.(\ref{eq: Psi_B}) as

	\begin{align}\label{eq: d_Psi}
			\ket{\partial_{B_0}\Psi_{\mathbf{B}}(t)} = \hat{\Lambda}_p \hat{\textsf{C}}_{\mathbf{B}}& \ket{\partial_{B_0} \Psi_{\mathbf{B}}(t-1)} \nonumber \\ +& \hat{\Lambda}_p(\partial_{B_0} \hat{\textsf{C}}_{\mathbf{B}})\ket{\Psi_{\mathbf{B}}(t-1)}.
	\end{align}

Sequentially substituting Eq.(\ref{eq: d_Psi}) in Eq.(\ref{eq: d_rho}), Eq.(\ref{eq: d_rho}) in Eq.(\ref{eq: SLD}), and Eq.(\ref{eq: SLD}) into Eq.(\ref{eq: QFI}), we can calculate the QFI provided by our system after $t$ time steps. 

Similarly, to calculate FI provided by the position (PFI) and spin (SFI) measurements, we use Eq.(\ref{eq: FI}), replacing the trace operator with partial trace over spin and position states, respectively. Let us first discuss how the theoretical maximum sensitivity of our magnetometer, provided by the QFI, is spread over the magnetic field strengths.

QFI increases as total time steps (T) increases irrespective of what initial spin state ($\ket{s}$) and direction ($\mathbf{\hat{n}}$) of the magnetic field is chosen. QFI is a periodic function with a period of $\pi$. The variables on which the evolution of DTQW depends are the initial spin state $\ket{s}$, the ratio of total time steps to the boundary point of the lattice ($T/a$), and the coin parameter ($\theta$). We now examine the effects of each variable on the QFI distributions, $H(\lambda)$.

\subsubsection{QFI of DTQW in presence of B${\hat{i}}$ ($H_x(\omega)$)}

Similar to their variance counterparts ($\sigma_{x}^2)$, points where QFI peaks can also be shifted when B$\mathbf{\hat{i}}$ is affecting the DTQW (see Fig.\ref{QFI}). To shift $H_x(\omega)$ by an amount ($\Delta \omega$), change the coin parameter ($\theta$) by the same value. Moreover, the maximum value of $H_x(\omega)$ gets doubled when $\ket{s}$ is changed from $\ket{+}$ to $\ket{0}$ or $\ket{1}$. Keeping the total time steps (T) fixed and reducing the boundary points of the lattice (a) less than a certain number, $N_0$, changes the nature of $H_x(\omega)$. This indicates that if we do not reduce $a$ less than $N_0$, we can get the same information about the unknown parameter, $B_0$. Decreasing $a$ below $N_0$ distorts $H_x(\omega)$ without raising its maximum value; hence, it is of no potential use.

\subsubsection{QFI of DTQW in presence of B${\hat{j}}$ ($H_y(\omega)$)}

When B$\mathbf{\hat{j}}$ is present, QFI ($H_y(\omega)$), similar to its variance cousin ($\sigma^2_y$), can only be shifted in steps of $\pi/2$ (Fig.\ref{QFI}). Apart from shifting $H_y(\omega)$ by a rigidly fixed amount, a change in $\theta$ can reduce the height of its peak but cannot increase it. At coin parameter, $\theta = \pi/4$, $H_y(\omega)$ becomes constant irrespective of whether $\ket{+}$ or $\ket{0}$ or $\ket{1}$ is used as the initial spin state. Again, decreasing boundary point $a$ up to a fixed number, $N_0$, does not change $H_y(\omega)$. When $a<N_0$, $H_y(\omega)$ gets distorted without increasing its maximum value.

\subsubsection{QFI of DTQW in presence of B${\hat{k}}$ ($H_z(\omega)$)}

The peaks of $H_z(\omega)$ are fairly rigid. We cannot shift them by changing the coin parameter $\theta$ (see Fig.\ref{QFI}). A change in the value of $\theta$ only changes the value of $H_z(\omega)$ \textemdash shifts the entire plot in the vertical direction with minor changes in their forms. For $\ket{s} = \ket{+}$ when $T/a=1$ and $\theta=0$, we get a surprisingly high and constant value of $H_z(\omega)$. For other values of $\theta$, when the spin state $\ket{+}$ is used, $H_z(\omega)$ takes sinusoidal form similar to $\sigma^2_{z}$. When we use unsymmetrical spin state $\ket{0}$ or $\ket{1}$, $H_z(\omega)$ becomes a constant function of $\omega$ for all values of $\theta$. Decreasing the boundary point, $a$, reduces the maximum value of $H_z(\omega)$ and distorts its form. 

The field magnitudes ($\omega = \gamma B_0 /2$) where QFI drops to minimum values are tough to estimate. This is because Fisher information obtained by any measurement will be less than QFI ($F(\omega) \le H(\omega)$). However, for fields in the \emph{x}-direction, owing to their \emph{shift} property, we can reliably estimate any magnitude by suitably shifting the peaks. For fields in \emph{y}-direction, however, the magnitudes where $H_y(\omega) $ drops to minima cannot be precisely estimated. 

\begin{figure}[h]
	\includegraphics[width=0.49\columnwidth]{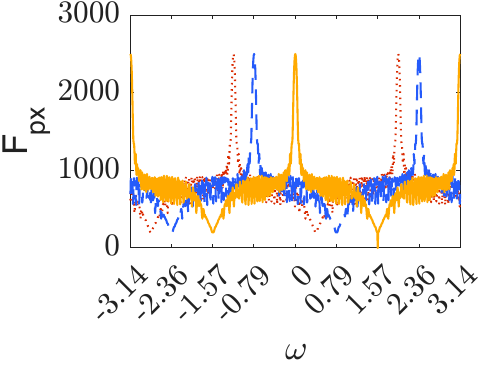}
	\includegraphics[width=0.49\columnwidth]{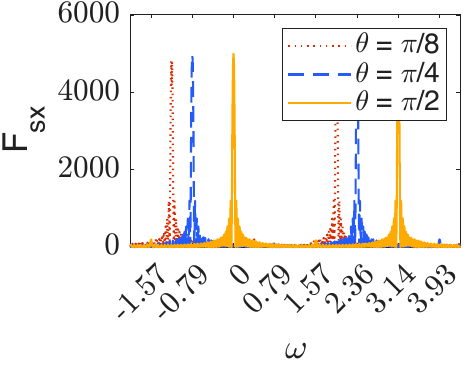}
	\includegraphics[width=0.49\columnwidth]{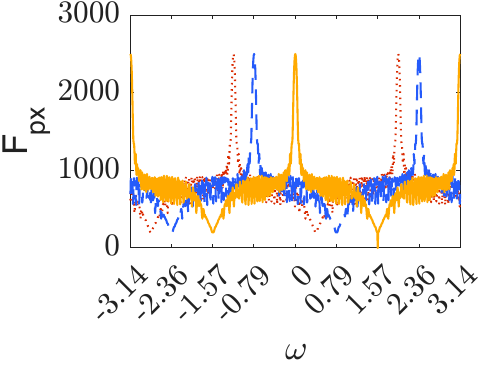}
	\includegraphics[width=0.49\columnwidth]{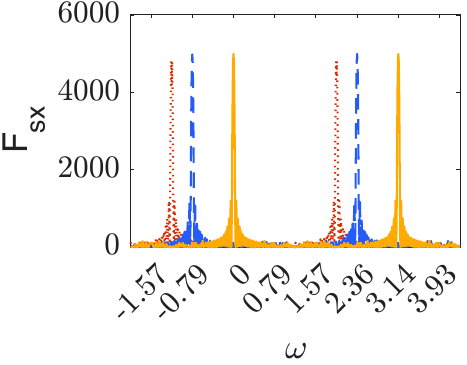}
	\includegraphics[width=0.49\columnwidth]{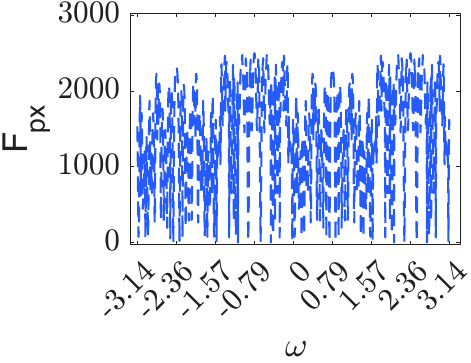}
	\includegraphics[width=0.49\columnwidth]{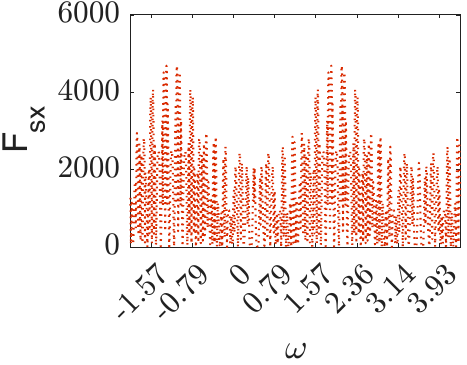}

	\caption{Fisher information provided by position measurements ($F_{px}$, left panel) and spin measurements ($F_{sx}$, right panel) of the particle undergoing DTQW with $T=50$ is plotted against different field magnitudes ($\omega$). A ratio of total time steps to the boundary of 1D lattice (T/a) equal to 1 is used for the two plots in the top panels. While $T/a = 2$ and $T/a = 50$ are used for the two plots in the middle and the two at the bottom, respectively. In all cases, the particle begins with initial spin state $\ket{1}$, magnetic fields in the \emph{x}-direction are applied to the system.}
	\label{FI}
\end{figure}

	\subsubsection{FI provided by the spin and the position measurements}

Having discussed the maximum theoretical efficiency of our system, we assess how well our system estimates the unknown parameter, $B_0$, by using the results of spin and position measurements. We eliminated fields in the \emph{y}-direction as the minima cannot be shifted; magnetic fields in the z-direction showed the highest QFI. However, neither the position FI nor the spin FI show peaks higher than $1\%$ of the peaks achieved by PFI and SFI for B$\mathbf{\hat{i}}$. Hence, detecting B$\mathbf{\hat{k}}$ is less fruitful than detecting B$\mathbf{\hat{i}}$ even though QFI provided by former is higher (in one case) than the latter. As a result, we will exclusively discuss the Fisher information provided by spin and position measurements when the magnetic field is in the \emph{x}-direction.

Figure \ref{FI} shows that PFI ($F_{px}(\omega)$), as expected, can be shifted by any desired amount by changing the coin parameter $\theta$. Again it comes to us with no surprise that using the initial spin state $\ket{0}$ or $\ket{1}$ turns out to give higher maximum value of $F_x(\omega)$ than when $\ket{+}$ is used (Considering $H_x(\omega)$ attained higher maximum value when $\ket{s} = \ket{0}$ or $\ket{1}$ was used). SFI, $F_{sx}(\omega)$, behaves in a similiar fashion. Its peaks can be shifted, $\ket{0}$ or $\ket{1}$ gives higher maximum than $\ket{+}$. However, the maximum value of $F_{sx}(\omega)$ is double of that of $F_{px}(\omega)$. In particular, $F^{max}_{sx} = 2 F^{max}_{px} = 2T^2$, for DTQW of $T$ time steps. The peaks are sharper and further apart in $F_{sx}(\omega)$ than $F_{px}(\omega)$ (see Fig.\ref{FI}). Thus, if we already know the possible magnitude of B$\mathbf{\hat{i}}$ (say through some theoretical calculation or prior measurements) and we need to verify the results with higher precision, we measure the spin of the walker at the end of the quantum walk. In contrast, when we want our system to be sensitive to a broader range of magnetic field strengths, position measurement is a better option.

Notice that similar to $H_x$, the nature of $F_{px}(\omega)$ and $F_{sx}(\omega)$ do not change if we keep total time steps ($T$) constant and reduce the boundary points of our lattice (a) upto a fixed amount $N_0$ ($N_0 < T$). In Figure \ref{FI}, we show this by plotting $F_{px}$ and $F_{sx}$ for $T/a=1$ (in the top panel) and $T/a=2$ (in the middle panel). The two sets of plots are identical even though the boundary of the lattice is reduced from $a=50$ to $a=25$, keeping the total time steps unchanged ($T =50$ in both cases). Estimating the parameter, $B_0$, with the same precision but using a smaller lattice size saves considerable space. We found that when DTQW with $T$ time steps is being used, it is sufficient to use a lattice with boundaries at $-T/2$ and $T/2$ to estimate the magnitude of a magnetic field without losing any precision. The reasons are discussed in the Appendix.

Finally, in Fig.\ref{FI/QFI}, we show the ratio $R=FI/QFI$ to assess the performance of our scheme for both position and spin measurements. For position measurements (top panel, Fig.\ref{FI/QFI}), we get peaks close to one for the field magnitudes where both $F_{px}$ and $H_x$ dropped to the minimum. Therefore, although these magnitudes seem to be optimally estimable, one should tune the peaks with $R_{px} = 0.5$ to be around those field magnitudes that need to be precisely estimated. Because at $\omega$ values where $R_{px} = 0.5$, both $F_{px}$ and $H_x$ are the highest. Contrarily, for spin measurements (bottom panel, Fig.\ref{FI/QFI}), the field magnitudes, where $F_{sx}$ and $H_x$ attain highest values, are also the points where their ratio $R_{sx} = 1$. Consequently, spin measurements provide optimum estimability of the field magnitudes where $R_{sx}$ peaks. Moreover, the freedom to shift the peaks to any desired field magnitude makes the scheme robust and tunable. 

\subsection{The minimum RMSE in the estimation of magnetic field}

As discussed, assuming an efficient unbiased estimator is available, the inequality in Eq.(\ref{eq: Cramer Rao}) saturates. Thus, we can calculate the RMSE using FI. Using the peak values of PFI ($F^{max}_{px}$) and SFI ($F^{max}_{sx}$) turning the inequality in Eq.(\ref{eq: Cramer Rao}) into an equality, the RMSE in estimating $\omega$ using position measurements takes the form:
\begin{equation} \label{eq: RMSE}
	\delta \omega = \frac{1}{T \sqrt{M}}.
\end{equation}

Note that $\delta \omega = \gamma (\delta B)/2$, and $\gamma = g_s \mu_B /\hbar$, where is $g_s$ is the gyromagnetic ratio, $\mu_B$ is the Bohr magneton. Say we use an electron with $g_s = -2.0023$ undergoing DTQW of total time steps 50 over a 1D lattice of size 51 ($a=25$). We align the system such that the magnetic field takes x-direction with respect to the system. Using a single position measurement ($M=1$) and an efficient unbiased estimator, we can estimate the magnitude of the field with the RMSE in the estimate of the order 0.1 picoTesla.

\begin{figure}[h]
	\includegraphics[width=0.9\columnwidth]{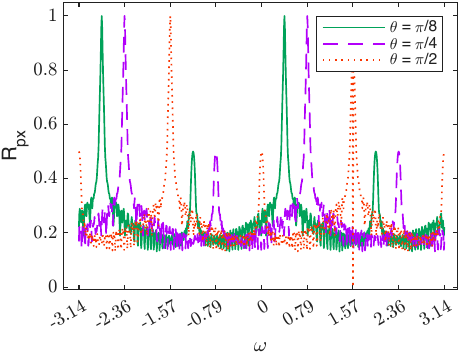}
	\includegraphics[width=0.9\columnwidth]{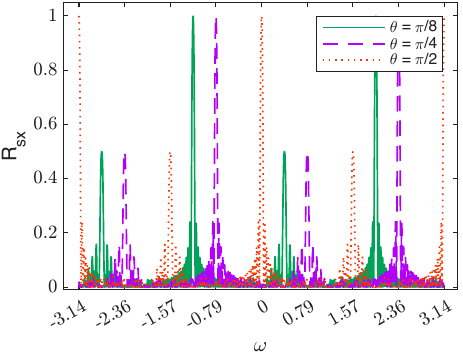}
	
	\caption{Ratio of the Fisher information provided by position measurement of the particle to the quantum Fisher information (QFI), $R_{px}$, is plotted against various magnitudes ($\omega$) of the magnetic fields (top). The ratio of spin Fisher information to QFI, $R_{sx}$, is plotted (bottom). In all cases, magnetic fields in the \emph{x}-direction are used. All plots are plotted after fifty steps of DTQW.}
	\label{FI/QFI}
\end{figure} 
		
	\section{Conclusions}

	In this work, we have studied the effects of homogeneous magnetic fields on a spin-half particle undergoing DTQW . The study supports  the  use of DTQW of the spin-half particle as a magnetometer for detecting and estimating magnetic fields. In our study we first showed that magnetic fields affect the multi-path interference of a particle undergoing DTQW. We used the variance of the position probability distributions to examine the effects of the magnetic fields. We found that variance is distributed periodically over the magnetic field's magnitude and shows different natures for fields in different directions. In particular, for magnetic fields in the \emph{x}-direction, the variance plots can be shifted by changing the parameter of the coin operator evolving the walk. We found that the direction of the magnetic field, maximally affecting the quantum walk, changes with the field magnitude. Furthermore, using the \emph{shift} property of magnetic fields in the \emph{x}-direction, by suitably changing the coin parameter, we can make the \emph{x}-direction the maximally affecting direction for all magnitudes. However, doing that required turning the coin operator to the identity operator which impedes the multi-path interference otherwise provided by DTQW and reduces the estimatability of the field magnitude.
	
	Therefore, we turned our attention to assessing how well our system can estimate the strengths of magnetic fields. We used QFI to study the maximum sensitivity of our system. We also examined whether position and spin measurements of the particle can be used to estimate the magnitude of the field. The plots of position and spin Fisher information (PFI and SFI, respectively) showed that this indeed is possible. The Fisher information plots showed peaks around some specific field magnitudes, suggesting that they are more efficiently estimatable than the other magnitudes. In addition, we observed the \emph{shift} property of fields in \emph{x}-direction for FI plots as well, which allows us to shift the peaks of the PFI and SFI plots to any desired magnitude of the magnetic field. This showed that our system, as a magnetometer, can be tuned to detect desired magnetic fields by simply changing the coin parameter used in the DTQW. Furthermore, high values of FI/QFI ratios indicated that carefully selecting the coin parameter $(\theta)$ maximizes the information extraction about the field's magnitude from the DTQW of the particle. In particular, when calculations were done for an electron undergoing DTQW of fifty time steps, i.e., using the system for fifty seconds, the RMSE in the estimate calculated using a single position measurement was of order of 0.1 picoTesla. Two leading candidates for detecting weak magnetic fields are magnetometers utilizing nitrogen-vacancy (NV) centers in diamond, boasting sensitivities of order of picoTeslas (15 pT/$\sqrt{\text{Hz}}$ \cite{xuRecentAdvancesApplications2023} for ensemble DC magnetometry), and superconducting quantum interference devices (SQUIDs), reaching sensitivities in the femtoTesla range (1--7 fT/$\sqrt{\text{Hz}}$) \cite{vettoliereHighlySensitiveTunable2023} at the cost of cryogenic cooling. The sensitivity of our scheme, contrastingly, depends on the time steps of DTQW and the number of measurements and can hence be further improved by using an ensemble of magnetometers (increasing M) for higher time intervals (increasing T) [Eq.(\ref{eq: RMSE})] without possibly needing cryogenic cooling.
	
	Detection is limited to static homogeneous magnetic fields. One way to overcome this shortcoming is that when a unidirectional inhomogeneous magnetic field is present, one can use several setups to detect the magnetic field strengths at different points. Reliable detection of the direction of magnetic fields is also a challenge. Moreover, Eq.(\ref{eq: RMSE}) clearly shows that using DTQW of a single fermion, we can hit the so-called standard quantum limit (SQL) having a $1/\sqrt{M}$ dependence on the number of measurements. However, this limit is beaten by the ultimate quantum limit of precision -- the Heisenberg limit \cite{giovannettiQuantumEnhancedMeasurementsBeating2004a}. Contrary to SQL, the Heisenberg limit has $1/M$ dependence and can be achieved by carefully choosing entangled states that exhibit strong correlations among the particles. Through this work, we achieved the SQL of precision in estimating magnetic fields by using quantum interference and entanglement of the spin and position of a fermion provided by DTQW dynamics. The entanglement between position and spin allowed us to effectively estimate magnetic fields' magnitudes by measuring the particle's position or spin. Meanwhile, future research ventures utilizing multi-fermion configurations, meticulously crafted entangled states, and tailored measurement protocols can be potentially explored to unlock the Heisenberg limit within the DTQW framework.
    If we can surmount the experimental challenges to make a suitably small setup, this magnetometer has the potential to be used in vast arenas wherever magnetic field strengths are required to be estimated with ultra-high precision.
		
	\section{Appendix}
\begin{figure}[t]
		\includegraphics[width=0.49\columnwidth]{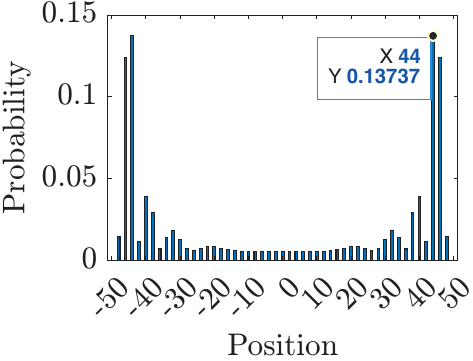}
		\includegraphics[width=0.49\columnwidth]{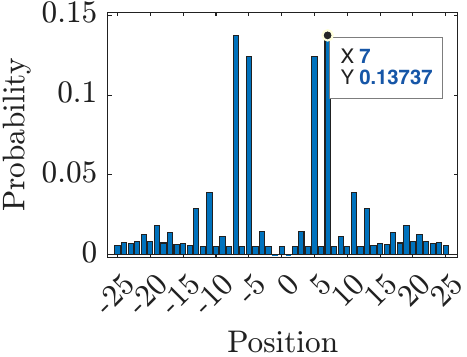}
		\includegraphics[width=0.49\columnwidth]{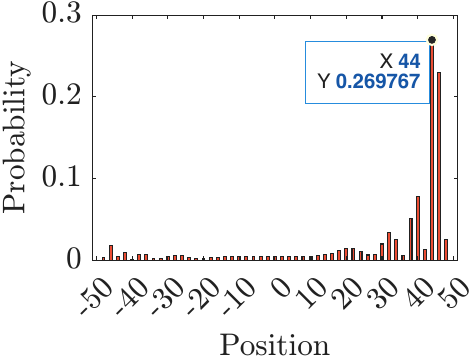}
		\includegraphics[width=0.49\columnwidth]{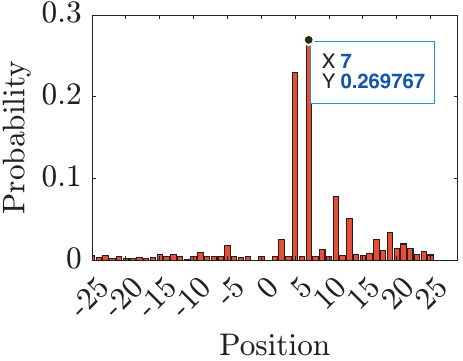}
		
		\caption{Position PD histograms for unbounded (left panels) and bounded (right panels) DTQW after 50 time steps. Initial spin state $\ket{+}$ is used for both plots in top panels while $\ket{1}$ is used for the rest. Position PD histograms of bounded DTQWs are obtained by folding that of the unbounded DTQWs about a verticle line at $x = \pm 25.5$ units along the \emph{x}-axis.}
		\label{histograms}
	\end{figure}

	We observed that the nature of QFI and PFI remained unchanged when we decreased the lattice boundaries, $a$, upto a certain number $N_0$. We will discuss the reasons in this section. In a bounded DTQW, the walker’s superposition terms hitting the bounds ($\pm a$) get reflected. This occurs due to the structure of the shift-operator [Eq.(\ref{eq: bounded_shift_operator})] evolving the bounded walk. From Fig.\ref{histograms}, we observe that the position PD histogram of the unbounded walk folds inwardly from verticle lines at $x=a+0.5$ and $x=-a-0.5$ to give the bounded walk position PD. Equivalently, the probability amplitudes of the bounded walk ($p_B (x)$) and the unbounded walk ($p_{\inf}(x))$ are related as $p_{\inf}(x = a + k) = p_{B} (x = a – (k-1))$ when all parameters are kept identicle for both kinds of walk. Figure \ref{histograms} further illustrates this: showing the probability values by putting $a=25$ and $k= 19$ for both variants. 
	In an unbounded DTQW with even $T$, the probability amplitudes at odd position spaces are zero. When we keep $a$ an odd number, the unbounded walk folds in the following manner: the probability values at even sites beyond the bounds ($\pm a$) goes to the odd sites between them. For DTQW with even $T$, the smallest safest bound that accomplishes this task is $T/2$. Consequently, the position PD of the bounded walk formed in this manner contains the same information present in that of the unbounded version (See Fig.\ref{FI}).

	\section{acknowledgments}		
	K.S. would like to thank Himanshu Sahu for meaningful discussions, Anshika Ranjan for her suggestions and his family for their emotional support.
	
	\bibliographystyle{apsrev4-2}
	\bibliography{QuantumMagnetometryDTQW_Revised}

\end{document}